\relax
\documentstyle [art12]{article}
\begin {document}
\bibliographystyle {plain}

\title{\bf Antiferromagnetic Spin-Ladders: Crossover Between Spin S =
1/2\\
 and S = 1 Chains.}
\author {D. G. Shelton,  A. A. Nersesyan$^+$ and A. M. Tsvelik}
\maketitle
\begin {verse}
$Department~ of~ Physics,~ University~ of~ Oxford,~ 1~ Keble~ Road,$
\\
$Oxford,~OX1~ 3NP,~ UK$\\
$and$\\
 $^+Institute~ of~ Physics,~
Georgian~ Academy~ of~ Sciences,$
\\
$Tamarashvili~ 6,~ 380077,~ Tbilisi, ~Georgia$
\end{verse}
\begin{abstract}
\par
We study a model of two weakly coupled isotropic
spin-1/2 Heisenberg chains with an antiferromagnetic coupling along
the chains (spin ladder). It is shown that the system always has a
spectral gap. For the case of identical chains the model in the
continuous limit is shown to be equivalent to four
decoupled  non-critical Ising
models with the Z$_2\times$SU(2)-symmetry. For this case
we obtain the  {\it exact} expressions for asymptotics of spin-spin
correlation functions.  It is  shown that when the chains have
different exchange
integrals $J_1 >> J_2$ the spectrum at low energies
is described by the
O(3)-nonlinear sigma model. We discuss the topological order parameter
related to the gap formation and give a detailed description of the
dynamical magnetic susceptibility.
\end{abstract}
cond-mat/9508047

\sloppy
\par
\section{Introduction}

 It has been widely recognized that one-dimensional antiferromagnets with
half-integer and integer spins have  dramatically different
excitation spectra. The original theoretical
prediction  by Haldane\cite{haldane} that Heisenberg chains with
half-integer spin are gapless, whereas those with integer spin are gapped,
has been confirmed experimentally \cite{s1}.
To gain insight into the physics underlying this beautiful
result, one may study systems intermediate between spin S = 1/2 and S
= 1. The
simplest of these is the ``Heisenberg
spin ladder", which has isotropic couplings $J_{||}$ along the chains and
$J_{\perp}$ between them. Although such  systems
with both
antiferromagnetic \cite{moreo},\cite{millis},\cite{dagotto}, \cite{rice},
\cite{white}
and
ferromagnetic \cite{hida},\cite{tokada},\cite{nomura},\cite{watanabe}
interchain
couplings
have  been the subject of
considerable recent theoretical interest, certain problems remain
unresolved leaving  room for our
contribution.

 In this paper we present our analysis of a weakly coupled spin ladder
$J_{||} >> |J_{\perp}|$.
We have found this limit more interesting
from the theoretical point of view for it allows us to study the
crossover between the gapless spin S = 1/2 regime and the
strong coupling limit corresponding to the S = 1 chain. Unfortunately, our
results have only qualitative validity for the
known experimental
realizations of double chain ladders ( Sr$_{n - 1}$Cu$_{n+1}$O$_{2n}$
\cite{strontium} and
 (VO)$_2$P$_2$O$_7$\cite{vanadium}) where both exchange integrals are
of the same order.

 The paper is organized as follows. In Section 2 we derive the
continuous version of the spin ladder Hamiltonian for the case of
identical chains. To achieve this we
employ the bosonization approach, but the resulting effective theory
is most simply represented in terms of fermions.
In this representation the effective Hamiltonian  of the spin ladder
contains  four species of weakly interacting real
fermions\footnote{The difference between ordinary (Dirac) and real
(Majorana) fermions is that the latter ones have only positive
energies $\epsilon(p) = \sqrt{p^2 + m^2}$. Therefore one can always
describe one Dirac fermion as a superposition of two Majorana
fermions.}.
Three of these modes comprise a degenerate triplet and the remaining
one lies above having a mass approximately three times as
big. The magnitude of the mass gaps is of the order of
the interchain exchange.
For any sign of the interchain coupling,
the  leading asymptotics of the correlation functions are
determined by the triplet of Majorana fermions as for the S = 1 chain
\cite{tsvelik}. The fact that the low energy sector of the model is
essentially a free theory makes it possible to obtain non-perturbative
expressions for asymptotics
of all correlation functions. This is done in Section 3. In Section 4
we discuss a situation where the ladder consists of
inequivalent chains. It is  shown that, in the limit when
exchange integrals on the chains strongly differ,
the low-lying
excitations  are described by the O(3)-nonlinear sigma model. The
adequacy of this treatment is guaranteed by the fact that this sigma
model has a small bare coupling constant. As we have pointed out,
the excitation spectrum of the spin ladder always has a gap. The appearence of
such a gap is usually related to some symmetry breaking process.
In Section 5 we discuss this process and derive the corresponding
order parameter which happens to be nonlocal
(the so-called string order parameter).
The paper has a Conclusion and two  Appendices where
we provide  technical details about
bosonization and string order parameters.

\section{Coupling of identical chains; the Abelian bosonization. }

 In this Section we apply the  Abelian bosonization method to
the spin-ladder model
\begin{equation}
H = J_{\parallel} \sum_{j=1,2} \sum_n {\bf S}_j (n) \cdot {\bf S}_j (n+1) +
J_{\perp} \sum_n {\bf S}_1 (n) \cdot {\bf S}_2 (n)
\end{equation}
describing two antiferromagnetic ($J_{\parallel} > 0$) spin-1/2
Heisenberg chains with a weak interchain coupling ($|J_{\perp}| \ll
J_{\parallel}$) of arbitrary sign.  Abelian bosonization is a well-known
procedure,
but for the sake of completeness we briefly overview it in the Appendix A.
In the continuum limit, the critical properties of isolated S = 1/2 Heisenberg
chains are described in
terms of massless Bose fields $\phi_j (x)~(j = 1,2)$:
\begin{equation}
H_0 = \frac{v_s}{2} \sum_{j=1,2}  \int dx [~ \Pi^2 _j (x) + (\partial_x \phi_j
(x))^2~]
\end{equation}
where the velocity $v_s \sim J_{\parallel} a_0$ and $\Pi_j$ are the momenta
conjugate to $\phi_j$.
The interchain coupling
\begin{equation}
H_{\perp} = J_{\perp} a_0 \int dx \left[ ~
{\bf J}_1 (x) \cdot {\bf J}_2 (x) + {\bf n}_1 (x) \cdot {\bf n}_2 (x) ~\right
]. \label{eq:interch}
\end{equation}
is expressed in terms of the operators ${\bf J}_j (x)$ and ${\bf n}_j (x)$
which represent, respectively, the slowly varying and staggered parts of local
spin density operator and are defined in the Appendix A.
According to (\ref{eq:scaldim}), the current-current term in
(\ref{eq:interch}) is marginal, while interaction of the staggered parts
of the spin densities is strongly relevant.
So we start our analysis by dropping the former term (its role will be
discussed later). Using then  bosonization formulas (\ref{eq:bosstag}) for
${\bf n}_j (x)$, we get
$$
H_{\perp} = \frac{J_{\perp} \lambda^2}{\pi^2 a_0} \int dx~
[  \frac{1}{2} \cos \sqrt{2 \pi} (\phi_1 + \phi_2)
+ \frac{1}{2} \cos \sqrt{2 \pi} (\phi_1 - \phi_2) $$
$$
+ \cos \sqrt{2 \pi} ({\theta}_1 - {\theta}_2)]
$$
where $\theta_j (x)$ is the field dual to $\phi_j (x)$.
Denote
\begin{equation}
m = \frac{J_{\perp} \lambda^2}{2 \pi}
\end{equation}
and introduce linear combinations of the fields $\phi_1$ and $\phi_2$:
\begin{equation}
\phi_{\pm} = \frac{\phi_1 \pm \phi_2}{\sqrt{2}} \label{+-defs}
\end{equation}
The  total ($\phi_+$) and relative ($\phi_-$) degrees of freedom decouple, and
the Hamiltonian of two identical Heisenberg chains transforms to a sum of
two independent contributions:
\begin{eqnarray}
H &=& H_+ + H_-  \label{eq:idenchains}  \\
H_+ (x) &=&  \frac{v_s}{2} \left( \Pi^2 _+  + (\partial_x \phi_+)^2 \right)
+ \frac{m}{\pi a_0} \cos \sqrt{4 \pi} \phi_+ \label{eq:+chan}  \\
H_- (x) &=&  \frac{v_s}{2} \left( \Pi^2 _-  + (\partial_x \phi_-)^2 \right)
+ \frac{m}{\pi a_0} \cos \sqrt{4 \pi} \phi_- +
\frac{2m}{\pi a_0} \cos \sqrt{4 \pi} {\theta}_- \label{eq:-chan}
\end{eqnarray}

 In the above derivation, the ${\bf J}_1 \cdot {\bf J}_2$-term has been
omitted as being only marginal, as opposed to the retained, relevant
${\bf n}_1 \cdot {\bf n}_2$-term.  It is worth mentioning that there are
modifications of the original
two-chain lattice model for which the  ${\bf J}_1 \cdot {\bf J}_2$-term
does not appear at all in the continuum limit, and mapping onto the model
(\ref{eq:idenchains}) becomes exact. In two such modifications,
the interchain coupling is changed to
\begin{equation}
H^{(A)}_{\perp} =
\frac{J_{\perp}}{2}  \sum_{n}
 {\bf S}_1 (n) \cdot   [ {\bf S}_2 (n) -  {\bf S}_2 (n + 1) ] \label{eq:A}
\end{equation}
or
\begin{equation}
H^{(B)}_{\perp} =
\frac{J_{\perp}}{4}  \sum_{n}
 [{\bf S}_1 (n) -  {\bf S}_1 (n + 1) ] \cdot
[ {\bf S}_2 (n) -  {\bf S}_2 (n + 1) ] \label{eq:B}
\end{equation}
The structure of these models explains why the low-energy physics
of two {\it weakly} coupled Heisenberg chains must not be sensitive
to the sign of the interchain coupling $J_{\perp}$. This conclusion is in
agreement with recent results of Ref.7.

Let us turn back to Eqs. (\ref{eq:+chan}) and (\ref{eq:-chan}).
 One immediately realizes that the critical dimension of all the cosine-terms
in
Eqs. (\ref{eq:+chan}), (\ref{eq:-chan}) is 1; therefore the model
(\ref{eq:idenchains}) is a theory of free massive fermions. The
Hamiltonian $H_+$ describes the  sine-Gordon model
at $\beta^2 = 4 \pi$; so it is equivalent to
a free massive Thirring model. Let us introduce a spinless
Dirac fermion  related to the scalar field $\phi_+$ via identification
\begin{equation}
\psi_{R,L} (x) \simeq (2 \pi a_0)^{-1/2}
\exp \left( \pm \mbox{i} \sqrt{4 \pi}~ \phi_{+; R,L} (x) \right)
\end{equation}
Using
$$
\frac{1}{\pi a_0} \cos \sqrt{4 \pi}~ \phi_{+} (x) =
\mbox{i} ~[~ \psi^{\dagger}_R (x) \psi_L (x) - h.c. ~]
$$
we get
\begin{equation}
H_+ (x) = - \mbox{i} v_s ( \psi^{\dagger}_R \partial_x \psi_R -
\psi^{\dagger}_L \partial_x \psi_L )
+ \mbox{i} m ( \psi^{\dagger}_R \psi_L - \psi^{\dagger}_L \psi_R )\label{mm}
\end{equation}
For future purposes, we introduce two real (Majorana) fermion  fields
\begin{equation}
\xi_{\nu} = \frac{\psi_{\nu} + \psi^{\dagger}_{\nu}}{\sqrt{2}}, ~~
\eta_{\nu} = \frac{\psi_{\nu} - \psi^{\dagger}_{\nu}}{\sqrt{2} i}, ~~
(\nu = R,L)
\end{equation}
to represent $H_+$ as a model of two degenerate massive Majorana fermions
\begin{equation}
H_+ = H_m [\xi] + H_m [\eta] \label{eq:M+}
\end{equation}
where
\begin{equation}
H_m [\xi] = - \frac{\mbox{i} v_s}{2} (\xi_R ~\partial_x \xi_R - \xi_L
{}~\partial_x \xi_L)
+ \mbox{i} m ~\xi_R \xi_L \label{eq:M+1}
\end{equation}
($ H_m [\eta]$ has the identical form).

 Now we shall demonstrate that the Hamiltonian
$H_-$ in (\ref{eq:-chan}) reduces to the Hamiltonian
of two {\it different} Majorana fields.
As before, we first introduce a spinless Dirac fermion
\begin{equation}
\chi_{R,L} (x) \simeq (2 \pi a_0)^{-1/2}
\exp \left( \pm \mbox{i} \sqrt{4 \pi}~ \phi_{-; R,L} (x) \right)
\end{equation}
\begin{eqnarray*}
\frac{1}{\pi a_0} \cos \sqrt{4 \pi}~ \phi_{-} (x) =
\mbox{i} ~[~ \chi^{\dagger}_R (x) \chi_L (x) - h.c. ], \\
\frac{1}{\pi a_0} \cos \sqrt{4 \pi}~ {\theta}_{-} (x) =
- \mbox{i} ~[~ \chi^{\dagger}_R (x) \chi^{\dagger}_L (x) - h.c. ]
\end{eqnarray*}
Apart from the usual mass bilinear term ("CDW" pairing), the
Hamiltonian $H_-$
also contains a "Cooper-pairing" term originating from the cosine of the dual
field:
\begin{eqnarray}
H_- (x) = - v_s ( \chi^{\dagger}_R \partial_x \chi_R -
\chi^{\dagger}_L \partial_x \chi_L ) \nonumber\\
+ \mbox{i} m ~( \chi^{\dagger}_R \chi_L - \chi^{\dagger}_L \chi_R )
+ 2 \mbox{i} m ~( \chi^{\dagger}_R \chi^{\dagger}_L - \chi_L \chi_R )
\label{m3m}
\end{eqnarray}
We introduce two Majorana fields
\begin{equation}
\zeta_{\nu} = \frac{\chi_{\nu} + \chi^{\dagger}_{\nu}}{\sqrt{2}}, ~~
\rho_{\nu} = \frac{\chi_{\nu} - \chi^{\dagger}_{\nu}}{\sqrt{2} i}, ~~
(\nu = R,L) \label{tr}
\end{equation}
The Hamiltonian $H_-$ then describes two massive Majorana fermions,
$\zeta_{R,L}$
and $\rho_{R,L}$, with masses $- m$ and $3m$, respectively:
\begin{equation}
H_{-} = H_{- m} [\zeta] + H_{3m} [\rho]
\end{equation}

Now we observe that
$\xi^1 \equiv \xi, ~\xi^2 \equiv \eta$ and $\xi^3 \equiv \zeta$
form a triplet of Majorana fields with the same modulus of mass $m$.
There is one more
field $\rho$ with a larger mass, $3m$. So,
\begin{equation}
H = H_{m} [\vec{\xi}] + H_{3m} [\rho]
\end{equation}
with
\begin{equation}
H_{m} [\vec{\xi}] =
\sum_{a=1,2,3} \{ - \frac{i v_s}{2}
(~\xi^a _R ~\partial_x \xi^a _R - \xi^a _L ~\partial_x \xi^a _L)
+ (-1)^{\delta_{a,3}}\mbox{i} m ~ \xi^a _R \xi^a _L \} \label{eq:S=1Maj}
\end{equation}

The exact degeneracy in the masses and, as a result, $O(3)$ invariance of the
Hamiltonian $H[\vec{\xi}]$ is recovered by a duality transformation
$\xi^3 _R \rightarrow - \xi^3 _R$ (with all the other Majorana fields
intact) that effectively changes the sign of the mass of the third Majorana
component. Let us denote the transformed triplet of fields by
$\bar{\xi}^a$
$$
\bar{\xi}^a _{\nu}
= (- 1)^{\delta_{a3}}~ (-1)^{\delta_{\nu, R}} \xi^a _{\nu}.
$$
The resulting model $H_m [\vec{\bar{\xi}}]$
was suggested as a
description of the $S = 1$ Heisenberg chain by Tsvelik
(\cite{tsvelik}). This equivalence follows from the fact that,
in the continuum limit,
the integrable S = 1 chain with the Hamiltonian
\begin{equation}
H = \sum_n[(\vec S_n\vec S_{n + 1}) - (\vec S_n\vec S_{n + 1})^2]
\end{equation}
is described by the critical Wess-Zumino model on the SU(2) group
at the level
$k = 2$, and the latter is in turn equavalent to the model of three massless
Majorana fermions, as follows from the comparison of conformal charges of
the corresponding theories:
$$
C^{WZW}_{SU(2),k=2} = \frac{3}{2} = 3 C_{Major.fermion}
$$
The $k = 2$ level, SU(2)
currents expressed in terms of the fields $\bar{\xi}^a$ are given by
\begin{equation}
I^a _{R,L} = - \frac{i}{2} \epsilon^{abc} ~\bar{\xi}^b _{R,L} \bar{\xi}^c
_{R,L}
\label{eq:k=2curr}
\end{equation}
When small deviations from criticality are considered, no single-ion anisotropy
($ \sim D (S^z)^2,~S=1$) is allowed to appear due to the original SU(2)
symmetry of the problem. So, the mass term in
(\ref{eq:S=1Maj}) turns out to be the only allowed relevant perturbation to the
critical $SU(2), k=2$ WZW model.

Thus, the fields $\xi^a$ describe triplet excitations related to the effective
spin-1 chain. Remarkably,
completely decoupled from them are singlet excitations described in terms of
the field $\rho$. Another feature is that this picture is valid for {\it any}
sign of $J_{\perp}$, in agreement with the effective lattice models
(\ref{eq:A}) and (\ref{eq:B}) which we actually are dealing with.

Since the spectrum of the system is massive, the role of
the so far neglected (marginal) part of the interchain coupling
(\ref{eq:interch}) is exhausted by renormalization of the masses and velocity.
Neglecting the latter effect, this interaction can be shown to
have the following invariant form
\begin{equation}
H_{marg} = \frac{1}{2} J_{\perp} a_0 \int dx~
[~ (I^a _R   I^a _L) + (\bar{\xi}^{a}_{R} \bar{\xi}^{a}_{L})~
(\rho_R \rho_L)~]
\end{equation}
which, after transforming back from $\bar{\xi}^a$ to $\xi^a$, reads
\begin{eqnarray}
H_{marg} = \frac{1}{2} J_{\perp} a_0 \int dx
[~ (\xi^1 _R \xi^1 _L)~(\xi^2 _R \xi^2 _L) -
(\xi^1 _R \xi^1 _L + \xi^2 _R \xi^2 _L)~(\xi^3 _R \xi^3 _L) \nonumber\\
+ (\xi^1 _R \xi^1 _L + \xi^2 _R \xi^2 _L - \xi^3 _R \xi^3 _L)
{}~(\rho_r \rho_L) ~] \label{eq:intmarg}
\end{eqnarray}
In a theory of N massive Majorana fermions, with masses $m_a~(a = 1,2,...,N)$
and a weak four-fermion interaction
$$
H_{int} = \frac{1}{2} \sum_{a \neq b} g_{ab} \int dx
(\xi^a _R \xi^a _L)~(\xi^b _R \xi^b _L),~~~~(g_{ab} = g_{ba})
$$
renormalized masses $\tilde{m}_a$ estimated in the first order in $g$ are given
by
\begin{equation}
\tilde{m}_a = m_a + \sum_{b (\neq a)} \frac{g_{ab}}{2 \pi v}~ m_b \ln
\frac{\Lambda}{|m_b|} \label{eq:massren}
\end{equation}
Using (\ref{eq:intmarg}) and (\ref{eq:massren}), we find renormalized values of
the masses of the triplet and singlet excitations:
\begin{eqnarray}
M_{trip} &=& m \left( 1 + \frac{5 J_{\perp} a_0}{4 \pi v} \ln
\frac{\Lambda}{|m|} \right) \\
M_{sing} &=& 3m \left( 1  +  \frac{J_{\perp} a_0}{4 \pi v} \ln
\frac{\Lambda}{|m|}\right)
\end{eqnarray}

\section{Correlation functions for the identical chains.}

 Since the singlet excitation with $m_s = 3m$ does not carry spin, its
operators do not contribute to the slow components of the total
magnetization. The latter is expressed in terms of the $k = 2$ $SU(2)$ currents
(\ref{eq:k=2curr}):
\begin{equation}
M^a \sim I^a _R + I^a _L.
\end{equation}
Therefore the two point correlation function of spin densities  at small wave
vectors $(|q| << \pi /a_0)$ is given by the simple fermionic loop. A simple
calculation
gives the following expression for its imaginary part:
\begin{equation}
\Im m\chi^{(R)}(\omega, q) = \frac{2q^2m^2}{s^3\sqrt{s^2 - 4 m^2}}\label{eq:f}
\end{equation}
where $s^2 = \omega^2 - v^2q^2$.
Thus the dynamical magnetic susceptibility at small wave vectors has a
threshold at $2m$.

 It turns out that it is possible to calculate exactly
the two-point correlation functions of the staggered magnetization.
This is due
to the fact that the corresponding operators of the Heisenberg chains
are related (in the continuum limit) to the order and disorder
parameter fields of 2d Ising models \cite{luther}, \cite{ogilvie};
the correlation functions of the
latter operators are known exactly even out of criticality \cite{sato}.

Using formulas (\ref{eq:bosstag}) of the Appendix A, the components of the
total (${\bf n}^{(+)} = {\bf n}_1 + {\bf n}_2$)
and relative (${\bf n}^{(-)} = {\bf n}_1 - {\bf n}_2$)
staggered magnetization can be represented as
\begin{eqnarray}
n^{(+)}_x \sim \cos \sqrt{\pi} \theta_{+} \cos \sqrt{\pi} \theta_{-},
&& n^{(-)}_x \sim \sin \sqrt{\pi} \theta_{+} \sin \sqrt{\pi} \theta_{-}
\nonumber\\
n^{(+)}_y \sim \sin \sqrt{\pi} \theta_{+} \cos \sqrt{\pi} \theta_{-},
&& n^{(-)}_y \sim \cos \sqrt{\pi} \theta_{+} \sin \sqrt{\pi} \theta_{-}
\nonumber\\
n^{(+)}_z \sim \cos \sqrt{\pi} \phi_{+} \cos \sqrt{\pi} \phi_{-},
&& n^{(-)}_z \sim \sin \sqrt{\pi} \phi_{+} \sin \sqrt{\pi} \phi_{-}
\label{eq:pmstag}
\end{eqnarray}
The fields $\phi _+, \theta _+$ and $\phi _-, \theta _-$
are  governed by the Hamiltonians
(\ref{eq:+chan}) and (\ref{eq:-chan}), respectively. Let us first consider
exponentials $\exp (\pm i \sqrt{\pi} \phi_+)$, $\exp(\pm i
\sqrt{\pi}\theta_+)$.
Their correlation functions have been
extensively studied in the context of the {\it noncritical} Ising
model (see, for example, Ref. \cite{sato}).
It has been  shown that these
bosonic exponents with scaling dimension 1/8 are expressed in
terms of the order ($\sigma$) and disorder ($\mu$) parameters of two Ising
models as follows:
\begin{eqnarray}
\cos(\sqrt\pi\phi_+) = \sigma_1\sigma_2, && \sin(\sqrt\pi\phi_+) = \mu_1\mu_2,
\nonumber\\
\cos(\sqrt\pi\theta_+) = \mu_1\sigma_2, && \sin(\sqrt\pi\theta_+) =
\sigma_1\mu_2 \label{eq:sig}
\end{eqnarray}
Let us briefly comment on this correspondence.

As already discussed, the $\beta^2 = 4 \pi$
sine-Gordon model $H _+$, Eq.(\ref{eq:+chan}), is equivalent to a model of two
degenerate massive Majorana fermions, Eqs.(\ref{eq:M+}), (\ref{eq:M+1}).
As is well known (see, {\it e.g.}, Ref. \cite{itz1}),
a theory of massive Majorana fermion describes long-distance properties of 2d
Ising model, the fermionic mass being proportional to
$m \sim t = (T - T_c)/T_c$. So, $H_+$ is equivalent to two decoupled 2d Ising
models. Let $ \sigma_j$ and $\mu_j$ $(j = 1,2)$ be the corresponding
order and disorder parameters.  At criticality (zero fermionic mass),
four products $\sigma_1 \sigma_2$,
$\mu_1 \mu_2$, $\sigma_1 \mu_2$ and $\mu_1 \sigma_2$
have the same critical dimension 1/8 as that of the bosonic exponentials $\exp
(\pm i \sqrt{\pi} \phi_+)$, $\exp(\pm i \sqrt{\pi}\theta_+)$.
Therefore there must be some correspondence between the two groups of four
operators which should also hold at small deviations from criticality.
To find this correspondence, notice that, as follows from (\ref{eq:+chan}), at
$m < 0~~<\cos \sqrt{\pi_+} \phi> \neq 0$, while $<\sin \sqrt{\pi} \phi_+> = 0$.
Since the case $m < 0$ corresponds to the ordered phase of the Ising systems
($ t < 0$), $<\sigma_1>~=~<\sigma_2> ~\neq 0$, while $<\mu_1>~=~<\mu_2> ~=~ 0$.
At $m > 0$ (disordered Ising systems, $t > 0$) the situation is inverted:
$<\cos \sqrt{\pi_+} \phi> \neq 0$, $<\sin \sqrt{\pi} \phi_+> \neq 0$,
$<\sigma_1>~=~<\sigma_2> = 0$,  $<\mu_1>~=~<\mu_2> \neq 0$. This explains
the first two formulas of Eq.(\ref{eq:sig}).

Clearly, the exponentials of the dual field $\theta_+$ must be expressed in
terms of
$\sigma_1 \mu_2$ and $\mu_1 \sigma_2$. To find the correct correspondence, one
has to take into account the fact that a local product of the order and
disorder operators of a single Ising model results in the Majorana fermion
operator, i.e.
$$
\xi^1 \sim \cos \sqrt{\pi} (\phi_+ + \theta_+) \sim \sigma_1 \mu_1,
{}~~
\xi^2 \sim \sin \sqrt{\pi} (\phi_+ + \theta_+) \sim \sigma_2 \mu_2.
$$
This leads to the last two formulas of Eq.(\ref{eq:sig}).

To derive similar expressions for the exponents of $\phi_-$ and $\theta_-$,
the following facts should be taken into account: (i) the Hamiltonian
(\ref{m3m}) describing $''-''$-modes
is diagonalized by the same transformation
(\ref{tr}) as the Hamiltonian (\ref{mm}) responsible for the
$''+''$-modes; (ii)
the Majorana fermions now have different masses, and (iii) one
fermionic branch has a negative mass. In order to take a proper
account of these facts one should recall the following.

(a) A negative mass means that we are below the
transition.

(b) It follows from (ii) that $''-''$ bosonic exponents are also expressed
in terms of order and disorder parameters of two Ising models,
the latter, however, being characterized by
different $t$'s. We denote these operators as $\sigma_3, \: \mu_3$
(mass $- m$) and $\sigma, \: \mu$ (mass $3m$).

(c) Operators corresponding to a  negative mass can be rewritten in
terms of the ones with the  positive mass using the Kramers-Wannier
duality
transformation
\begin{equation}
t \rightarrow - t, \: \sigma \rightarrow \mu, \: \mu \rightarrow
\sigma \label{dualx}
\end{equation}
Taking these facts into account we get the following expressions for
the  $''-''$ -bosonic exponents:
\begin{eqnarray}
\cos(\sqrt\pi\phi_-) = \mu_3\sigma, \: \sin(\sqrt\pi\phi_-) = \sigma_3\mu, \:
\cos(\sqrt\pi\theta_-) = \sigma_3\sigma, \: \sin(\sqrt\pi\theta_-) =
\mu_3\mu \label{eq:siga}
\end{eqnarray}

Combining Eqs. (\ref{eq:sig}) and (\ref{eq:siga}), from (\ref{eq:pmstag})
we get the following, manifestly SU(2) invariant, expressions:
\begin{eqnarray}
 n^{+} _x \sim \mu_1 \sigma_2 \sigma_3 \sigma,~~
n^{+} _y \sim \sigma_1 \mu_2 \sigma_3 \sigma,~~
n^{+} _z \sim \sigma_1 \sigma_2 \mu_3 \sigma \\
\label{eq:pstag}
n^{-} _x \sim \sigma_1 \mu_2 \mu_3 \mu,~~
n^{-} _y \sim \mu_1 \sigma_2 \mu_3 \mu,~~
n^{-} _z \sim \mu_1 \mu_2 \sigma_3 \mu
\label{eq:mstag}
\end{eqnarray}
It is instructive to
compare them with two possible representations
for the staggered
magnetization operators for  the S = 1 Heisenberg chain (\cite{tsvelik}):
\begin{equation}
S^x \sim \mu_1 \sigma_2 \sigma_3,~~
S^y \sim \sigma_1 \mu_2 \sigma_3,~~
S^z \sim \sigma_1 \sigma_2 \mu_3 \label{eq:S1stag1}
\end{equation}
or
\begin{equation}
S^x \sim \sigma_1 \mu_2 \mu_3,~~
S^y \sim \mu_1 \sigma_2 \mu_3,~~
S^z \sim \mu_1 \mu_2 \sigma_3 \label{eq:S1stag2}
\end{equation}
Agreement is achieved
if the singlet excitation band is formally shifted to infinity.
This implies substitutions $\sigma \simeq \langle \sigma \rangle
\neq 0$,
$\mu \simeq \langle \mu \rangle \simeq 0$ for ferromagnetic interchain coupling
($m \sim J_{\perp} < 0$), or $\sigma \simeq \langle \sigma \rangle \simeq 0$,
$\mu \simeq \langle \mu \rangle \neq 0$ for antiferromagnetic interchain
coupling ($m \sim J_{\perp} > 0$). Thus, we observe that, as expected,
for ferromagnetic (antiferromagnetic) interchain interaction
the staggered S=1 magnetization is determined by the total (relative) staggered
magnetization of the two-chain system.

 A more precise meaning of this approximation becomes apparent when  one
considers
asymptotic behaviour of the corresponding
two-point correlation functions in the two limits $r\rightarrow 0$ and
$r\rightarrow \infty$(\cite{mccoy}).  In the limit $r \rightarrow
\infty$ they are as
follows;
\begin{eqnarray}
\langle\sigma_a(r)\sigma_a(0)\rangle  = G_{\sigma}(\tilde r)
= \frac{A_1}{\pi}K_0(\tilde r) + O(e^{- 3\tilde r})\\
\langle\mu_a(r)\mu_a(0)\rangle = G_{\mu}(\tilde r)\nonumber\\
= A_1\left\{1 +
\frac{1}{\pi^2}\left[{\tilde r}^2\left(K_1^2(\tilde r) - K_0^2(\tilde r)\right)
-
\tilde rK_0(\tilde r)K_1(\tilde r) + \frac{1}{2}K_0^2(\tilde r)\right]\right\}
+ O(e^{- 4\tilde r})
\end{eqnarray}
where $\tilde r = rM$ (M = $m$ or $3m$), $A_1$ is a nonuniversal
parameter, and it has been assumed that $M$ is positive. If $M$ is
negative the correlation functions are obtained by simply
interchanging $\sigma$ and $\mu$, and putting $M\rightarrow -M$ (the
duality transformation (\ref{dualx})).
Therefore, as might be expected, at large distances, a difference
between the
 ladder and
the S = 1 chains appears  only in $\exp(- 3mr)$-terms due to the
contribution of  the
excitation branch with $M = 3m$ absent in the S = 1 chain.

In the limit $\tilde r \rightarrow 0$ the correlation functions are of
power law form:
\begin{equation}
 G_{\sigma}(\tilde r) =  G_{\mu}(\tilde r) = {A_2\over {\tilde r}^{1\over 4}}
\end{equation}
plus non-singular terms. The ratio of the constants $A_1$ and $A_2$ is a
universal quantity involving Glaisher's constant (A):
\begin{eqnarray}
{A_2\over A_1}&=&2^{- 1/6}A^{-3}\exp{1\over 4} \\
A&=&1.282427129...
\end{eqnarray}

We conclude this Section by writing down  the exact expression for the
staggered magnetization two-point correlation functions. The
correlation function for spins on the same chain is given by
\begin{eqnarray}
\langle n_1^a(\tau, x) n_1^a(0,0)\rangle = G_{\sigma}^2(mr)G_{\mu}(mr)
G_{\sigma}(3mr) + G_{\mu}^2(mr)G_{\sigma}(mr)
G_{\mu}(3mr)
\end{eqnarray}
The interesting asymptotics are
\begin{eqnarray}
\langle n_1^a(\tau, x) n_1^a(0,0)\rangle = \frac{1}{2\pi r}\tilde Z
 \: \mbox{at} \: mr << 1 \\
\frac{m}{\pi^2}ZK_0(mr)\{1 + \frac{2}{\pi^2}[(mr)^2(K_1^2(mr) -
K_0^2(mr)) - mrK_0(mr)K_1(mr) \nonumber\\
+ \frac{1}{2}K_0^2(mr)]\} + O(e^{-5mr});
\:  mr >> 1 \label{as}
\end{eqnarray}
where $r^2 = \tau^2 + v^2 x^2$ and
\begin{equation}
\frac{\tilde Z}{Z} = \frac{2^{4/3}\mbox{e}}{3^{1/4}}A^{-12} \approx 0.264
\end{equation}
 The complete expressions for the
functions $G_{\sigma, \mu}(\tilde r)$ are given in Ref.(\cite{mccoy}).
For the interchain correlation function we get
\begin{eqnarray}
\langle n_1^a(\tau, x) n_2^a(0,0)\rangle = G_{\sigma}^2(mr)G_{\mu}(mr)
G_{\sigma}(3mr) - G_{\mu}^2(mr)G_{\sigma}(mr)G_{\mu}(3mr)
\end{eqnarray}
At $mr << 1$ it decays as $(mr)^{-2}$; the leading asymptotics at $mr
>> 1$ is the same as (\ref{as}) (up to the -1 factor). The difference
appears only in terms of order of $\exp(- 5mr)$.
The important point is that at $mr >> 1$ the contribution from the
singlet excitation appears only in the fifth order in
$\exp(-mr)$. Therefore it is
unobservable by neutron scattering at energies below $5m$.

 Using the above expressions we can calculate the imaginary part of
the dynamical spin susceptibility in two different regimes. For $|\pi
- q| << 1$ we have
\begin{eqnarray}
\Im m\chi^{(R)}(\omega, \pi - q; q_{\perp}) = Z\left\{
\begin{array}{lr}
2\cos q_{\perp}[\frac{m}{|\omega|}\delta(\omega - \sqrt{v^2q^2 + m^2}) +
F(\omega,q)]& \omega < 5m\\
(1 + \cos q_{\perp})\frac{0.264}{\sqrt{\omega^2 - v^2q^2}} & \omega >> 5m
\end{array}
\right.
\end{eqnarray}
where the transverse "momentum" $q_{\perp}$ takes values $0$ and $\pi$.
The factor $Z$ is assumed to be $m$-independent so that at $m
\rightarrow 0$ we reproduce the susceptibility of non-interacting
chains. We have calculated the function  $F(\omega,q)$ only near the $3m$
threshold where it is equal to
\begin{equation}
F(\omega, q) \approx \frac{144}{\pi m^2}\sqrt{\omega^2 - v^2q^2 - 9m^2}
\end{equation}
For $|q| << 1$ we have
\begin{eqnarray}
\Im m\chi^{(R)}(\omega, q; q_{\perp}) = [1 + \cos^2(q_{\perp}/2)]f(s,
m)
\end{eqnarray}
where $f(s, m)$ is given by Eq.(\ref{eq:f}).

\section{Inequivalent chains; non-Abelian bosonization.}

 In this section we consider two interacting spin S = 1/2 chains with
different intrachain exchange integrals $J^1_{||} \neq J^2_{||}$. It turns out
that the most adequate approach in this case is non-Abelian
bosonization. The reason for this is that non-Abelian bosonization
explicitly preserves the SU(2) symmetry present in the
Hamiltonian. The Abelian bosonization approach which does not respect
this symmetry encounters difficulties.

 As shown by Affleck \cite{aff}, by a
mapping from a fermionic theory, the $S={1 \over 2}$ Heisenberg
antiferromagnet can be described  by a $k=1$, $SU(2)$
Wess-Zumino-Witten (WZW) model
with the following action;
\begin{eqnarray}
S_{k}&=&kW({\bf g}) \nonumber \\
W({\bf g})&=&{1 \over 16\pi}\int
Tr\left(\partial_{\mu}{\bf g}^{+}\partial_{\mu}{\bf g}\right) d^{2}x +\Gamma
\left( {\bf g} \right) \nonumber \\
\Gamma \left( {\bf g} \right) &=& {\mbox{i} \over 24\pi}\int d^{3}X
\epsilon^{\alpha \beta
\gamma}Tr\left({\bf g}^{+}\partial_{\alpha}{\bf
gg}^{+}\partial_{\beta}{\bf gg}^{+}\partial_{\gamma}{\bf g}\right)
\end{eqnarray}
where matrix ${\bf g}\in SU(2)$. There can in general be marginally
irrelevant perturbations to this theory, which  generate
logarithmic corrections to the correlation function exponents, but
do not change their qualitative behaviour (i.e. power law).
In general this model describes  not just the spin S = 1/2
Heisenberg chain, but any
(1+1)-dimensional  system of fermions with the
charge degree of freedom frozen out and no
gap in the spin sector.

 The WZW model may look unfamiliar, but it is not so difficult to deal with
since its  operators and their exact correlation
functions are already known from the
application of conformal field theory \cite{Knizhnik}(see also \cite{itz2}).
As we have mentioned above, the great advantage of the WZW model is that it
explicitly possesses the $SU(2)\times SU(2)$ symmetry of the
massless fermion spin sector, and is also critical (massless). In
(1+1)-dimensions
the distinction between relativistic fermions and bosons is illusory;
one can choose to think about a system in either representation (this
has been known for some time; hence the ``Luttinger liquid''). The
WZW model is therefore just a way of thinking about the
spin sector in terms of bosons; just as in Abelian bosonization, one
can represent operators from the fermionic theory in terms of those of
the bosonic theory and vice-versa. From a practical point of view, these
relations between the two sets
of operators can be thought of as readymade tools.
It is not necessary to worry about their slightly exotic appearance or their
justification in order to apply them. (But those seeking a deeper
appreciation are referred to the papers cited above.)

 The bosonized expression for the spin operator of the Heisenberg
chain is given by \cite{aff};
\begin{equation}
{\bf \overrightarrow{S}}_n =
{\bf \overrightarrow{J}}_R+{\bf \overrightarrow{J}}_L+const
(-1)^n Tr({\bf g}^+ {\bf \overrightarrow{\sigma}}
-{\bf g} {\bf \overrightarrow{\sigma}})
\end{equation}
where the currents are given by;
\begin{eqnarray}
J_R^{a}&=&-{\mbox{i} \over
2\pi}\mbox{Tr}(\partial_{-}{\bf g}){\bf g}^{+}T^{a} \nonumber \\
J_L^{a}&=&{\mbox{i} \over 2\pi}\mbox{Tr}{\bf g}^{+}\partial_{+}{\bf g}T^{a}
\end{eqnarray}
($T^a$ are the Pauli matrices - generators of the SU(2) group). These
currents satisfy the SU(2) Kac-Moody algebra described in the Appendix
A.

 Consider two Heisenberg chains coupled by an
antiferromagnetic nearest neighbour interaction. It can be represented
like this;
\begin{eqnarray}
S = W_1({\bf h})+W_2({\bf g}) \nonumber\\
+ \lambda_{1}\left[ {\bf \overrightarrow{H}}_R + {\bf
\overrightarrow{H}}_L\right] \left[ {\bf \overrightarrow{G}}_R+{\bf
\overrightarrow{G}}_L\right]
+\lambda_2 Tr\left[\left( {\bf g}-{\bf
g}^+\right)\overrightarrow{{\bf\sigma}}\right]
Tr\left[\left( {\bf h}-{\bf h}^+\right)\overrightarrow{{\bf\sigma}}\right]
\end{eqnarray}
where the dynamics of one chain is represented by the matrix ${\bf g}$ and
the currents ${\bf \overrightarrow{G}}_{R,L}$ and the other by ${\bf h}$ and
$ {\bf \overrightarrow{H}}_{R,L}$. The indices 1,2 distinguish between
different spin wave velocities. Without a loss of generality we can
put $v_1 > v_2$.

 The currents have conformal dimensions $(1,0)$ and $(0,1)$;  using
the formula for the conformal
dimensions of the matrices for the $SU(n)$ group
derived in Ref. \cite{Knizhnik}:
\begin{equation}
\Delta = {n^2-1 \over 2n(n+k)}
\end{equation}
we get that for $n = 2,\: k = 1$, ${\bf g}$ and ${\bf h}$ both have conformal
dimensions
$({1 \over 4}, {1
\over 4})$. The $\lambda_2$ term is therefore the relevant
interaction, whereas the current couplings are only marginal.
For this reason, the current interaction will be neglected at this
stage.
Then  the interaction can be written as;
\begin{eqnarray}
\mbox{Tr}[{(\bf g - \bf g^+)\overrightarrow{\sigma}}]\cdot \mbox{Tr}[{(\bf h -
\bf h^+)
\overrightarrow{\sigma}}] \nonumber\\
= {1 \over 2 }\left\{
\mbox{Tr}[{(\bf g - \bf g^+)(\bf h - \bf h^+)}] - \mbox{Tr}[{(\bf g - \bf g^+)
}]\mbox{Tr}[(\bf h - \bf h^+)]\right\}
\end{eqnarray}
Making the substitution ${\bf \alpha}={\bf gh^+}$, which leaves the
measure invariant, and
using the remarkable identity \cite{Knizhnik}:

\begin{equation}
W({\bf \alpha h^+})=W({\bf \alpha} ) + W({\bf h}) + {1\over 2\pi}\int
\mbox{Tr}{\bf \alpha}^+\partial_{-}{\bf \alpha h}^+\partial_{+}{\bf h} d^2 x
\label{knid}
\end{equation}
we arrive at the following expression for the action:
\begin{eqnarray}
S=[W_1({\bf h}) + W_2({\bf h})] \nonumber\\
+{1\over 2\pi}\int
\mbox{Tr}{\bf \alpha}^+\partial_{-}{\bf \alpha h}^+\partial_{+}{\bf h} d^2 x
\nonumber\\
+ W_1({\bf \alpha} )+ \lambda_2 \left[\mbox{Tr}({\bf\alpha} + {\bf\alpha^+}) -
\mbox{Tr}({\bf \alpha^+}{\bf h^+}^2 + H.c.) + \mbox{Tr}({\bf h^+} - {\bf h})
\mbox{Tr}({\bf
h^+}{\bf \alpha^+} - {\bf \alpha}{\bf h}) \right]
\end{eqnarray}
(here $\partial_{\pm} = \frac{1}{2}(\partial_{\tau} \mp
\mbox{i}\partial_x)$).

The identity (\ref{knid}) is nothing very mysterious. It is simply a
generalisation of an identity familiar from Abelian bosonization. To
see this, consider substituting explicitly for the special case of
Abelian bosonization, the U(1) fields
$e^{\mbox{i}\beta\phi_1}$ and $e^{\mbox{i}\beta\phi_2}$ for the matrices ${\bf
\alpha}$ and ${\bf h}$ respectively. Then the  WZW action, $W({\bf \alpha})$
reduces to the action for free scalar bosons as we would expect;

\begin{equation}
W\left( {\bf \alpha}=e^{\mbox{i}\beta\phi_1}\right) = {\beta^2 \over 4\pi}\int
\partial_+ \phi_1 \partial_- \phi_1 d^2 x ={\beta^2 \over 16\pi}\int
\left[ (\partial_x \phi_1)^2+ (\partial_{\tau} \phi_1)^2\right] d^2 x
\end{equation}
and the interaction term in (\ref{knid}) becomes;
\begin{equation}
{1\over 2\pi}\int
\mbox{Tr}{\bf \alpha}^+\partial_{-}{\bf \alpha h}^+\partial_{+}{\bf h} d^2 x
={-\beta^2 \over 4\pi}\int \left( \partial_+ \phi_1 \partial_- \phi_2
+ \partial_+ \phi_2 \partial_- \phi_1 \right) d^2 x
\end{equation}
The field ${\bf \alpha h^+}$ is $e^{\mbox{i}\beta
(\phi_1-\phi_2)}=e^{\mbox{i}\beta
\phi_-}$, and so the identity (\ref{knid}) becomes;
\begin{eqnarray}
 {\beta^2 \over 4\pi}\int
\partial_+ \phi_- \partial_- \phi_- d^2 x = &{\beta^2 \over
4\pi}&\left[  \int
\partial_+ \phi_1 \partial_- \phi_1 d^2 x + \int
\partial_+ \phi_2 \partial_- \phi_2 d^2 x \right.\nonumber\\
&-&\left. \int \left( \partial_+ \phi_1 \partial_- \phi_2
+ \partial_+ \phi_2 \partial_- \phi_1 \right) d^2 x \right]
\end{eqnarray}
Therefore the identity (\ref{knid}) is just an analogue  of the
following simple statement:
\begin{equation}
(\nabla (\phi_1-\phi_2))^2= (\nabla \phi_1)^2 +(\nabla \phi_2)^2
-2\nabla \phi_1 \cdot \nabla\phi_2
\end{equation}
where the last term is the ``interaction term''.

 We shall consider the most relevant
interaction, $\mbox{Tr}({\bf \alpha} + {\bf \alpha^+} )$ first. The effective
action for ${\bf
\alpha}$ is in this approximation;
\begin{equation}
S=W_1({\bf \alpha})+\lambda \mbox{Tr}({\bf \alpha} + {\bf \alpha^+})
\label{alpha1}
\end{equation}
 From the first order RG equation we get
\begin{equation}
{d\lambda \over d\ln L}\simeq (2-{1 \over 2})\lambda
\end{equation}
Integrating up to a scale where the coupling becomes of order 1 and
taking this to give some estimate of the dynamically generated mass,
one gets $M \sim \lambda^{2\over 3}$. Much more information can be
found by realising that the model (\ref{alpha1}) is equivalent to the
$\beta^2 =2\pi$ sine-Gordon model (see for example \cite{aff}).

 Thus on the scale $|x| >> M^{-1}$ the fluctuations of the
$\alpha$-field are frozen and we can approximate
\begin{equation}
\mbox{Tr}({\bf \alpha h}^+)Tr({\bf h}) \approx \langle \mbox{Tr}\alpha\rangle
:[\mbox{Tr}({\bf h})]^2:
\end{equation}
At this large scale the cross term containing
derivatives of $h$ and $\alpha$ gives the irrelevant contribution
\begin{equation}
S_{int} \sim M^{-2}\partial_{+}\partial_{-}{\bf
h}^+\partial_{+}\partial_{-}{\bf h}
\end{equation}

 Therefore the asymptotic behaviour at large distances
is governed by the following
action:
\begin{eqnarray}
S = W_1(h) + W_2(h) + c_2:[\mbox{Tr}({\bf h})]^2:
\end{eqnarray}
where $c_2 \sim \lambda^{4/3}$ and which can be further modified by the
coordinate rescaling:
\begin{equation}
x_0 =  \sqrt{v_1v_2}\tau, \: x_1 = x
\end{equation}
such  that we finally have
\begin{eqnarray}
S = S_0 + S_1 \label{wzw}\\
S_0 = \frac{1}{2c_1}\int \mbox{d}^2x\mbox{Tr}\left(\partial_{\mu}{\bf
h}^{+}\partial_{\mu}{\bf h}\right) d^{2}x +2\Gamma
\left( {\bf h} \right)\\
S_1 =  \int \mbox{d}^2x \tilde c_2\{:[\mbox{Tr}{\bf h^2}]: + :[\mbox{Tr}{\bf
(h^+)^2}]: - :[\mbox{Tr}({\bf h} - {\bf h^+})]^2:\} \label{eff}
\end{eqnarray}
where
\[
1/c = \sqrt{v_1/v_2} + \sqrt{v_2/v_1}
\]
 The model with action (\ref{wzw}) is not critical; coupling
constants $c_1, c_2$ undergo further renormalization. Let us show that
the coupling $c_2$ renormalizes faster to strong coupling.
To show this we shall suppose
that this is the case and check that the obtained result is
self-consistent. It is easy to check that the effective potential
(\ref{eff}) vanishes if ${\bf h}$ is a traceless matrix and has a
fixed determinant:
\begin{equation}
{\bf h} \approx  \mbox{i}(\vec{\bf \sigma}\vec
n), \: \vec n^2 = 1 \label{tras}
\end{equation}

Excitations, which correspond to configurations where Tr$\bf h \neq 0$,
acquire a gap.  The estimate for this gap  is
\begin{equation}
M_0^2 \sim c\sqrt{v_1v_2}\:c_2 \sim \min(v_1,v_2)\lambda^{4/3} \sim
\frac{v_2}{v_1}M^2
\end{equation}
On energies smaller than the gap one can treat the ${\bf h}$-matrix
as traceless.
Substituting expression (\ref{tras}) into Eq.(\ref{wzw}) we get the
O(3)-nonlinear sigma model as an effective action for small energies:
\begin{eqnarray}
S = \frac{1}{2\tilde c}\int \mbox{d}^2x(\partial_{\mu}\vec n)^2, \:
\vec n^2 = 1 \label{nf}\\
\frac{1}{\tilde c} = (\sqrt{v_1/v_2} + \sqrt{v_2/v_1})(1 - \langle
n_0^2\rangle)
\end{eqnarray}
The reason why the Wess-Zumino term effectively disappears from the
action is the following. After substituting Eq.(\ref{tras}) into the
expression for $\Gamma({\bf h})$ the Wess-Zumino term reduces to the
topological term:
\begin{equation}
2\Gamma(\mbox{i}\vec\sigma\vec n) = \frac{\mbox{i}}{4}\int d^2x
\epsilon_{\mu\nu}\left(\vec n[\partial_{\mu}\vec
n\times\partial_{\nu}\vec n]\right) = 2\pi\mbox{i}k
\end{equation}
where $k$ is an integer number. The factor in front of the topological
term is such that its contribution to the action is always a factor of
$2\pi$i and therefore does not affect the partition function.
The mass gap of the  model (\ref{nf}) is given by
\begin{equation}
M = M_0\tilde c^{-1}\exp(- 2\pi/\tilde c) \approx M[1 -
\frac{c}{2\pi}\ln(M/M_0)]\exp(-2\pi/c)
\end{equation}
As long as this gap is much smaller than $M_0$, the adopted
approach  is
self-consistent. The latter is achieved for any appreciable difference
between the velocities.

 Excitations of the O(3)-nonlinear sigma model are S = 1 triplets
(\cite{zamolodchikov}). Thus, the spectrum is qualitatively the same as for
identical chains. That is what one might expect because the model of
Majorana fermions is a strong coupling limit of the O(3)-nonlinear
sigma model (see Ref. \cite{tsvelik}).

 The correlation functions of the O(3)-nonlinear sigma model are known
only in the form of the Lehmann expansion (\cite{Smirnov}).
\begin{equation}
\langle \vec n(\tau,x) \vec n(0,0)\rangle \sim  K_0(mr) + O(\exp(- 3mr))
\end{equation}
Note that the first term in the expansion coincides with the one for
identical chains. Therefore a difference in dynamical magnetic
susceptibilities for both cases will become manifest only at energies
$\omega > 3m$. The lowest feature in $\Im m\chi^{(R)}(\omega, q)$ is
in both cases the sharp peak
\begin{equation}
\Im m\chi^{(R)}(\omega, q) \sim \frac{m}{\sqrt{q^2 +
m^2}}\delta(\omega - \sqrt{q^2 +
m^2})
\end{equation}
corresponding to the triplet excitation. Such a peak has been observed
in (VO)$_2$P$_2$O$_7$\cite{vanadium}.

\section{String order parameter in the spin-ladder model}

Den Nijs and Rommelse \cite{denNijs}
(see also \cite{girvin})
have argued that
the gapful Haldane phase of the S = 1 spin chain is characterized by a
topological order measured by the string order parameter
\begin{equation}
< O^{\alpha}> = \lim_{|n - m| \rightarrow \infty}
<~S^{\alpha}_{n} \exp ( \mbox{i} \pi \sum_{j=n+1}^{m-1} S^{\alpha}_{j})
S^{\alpha}_{m}~>,~~(S = 1, ~~\alpha = x,y,z ) \label{eq:strS=1}
\end{equation}
The nonzero value of $< O^{\alpha}>$ has been related to the breakdown
of a hidden $Z_2 \times Z_2$ symmetry \cite{Kohmoto}. In this section
we use the Abelian bosonization method (section 2) to construct the
string operator in the continuum  limit of the S = 1/2 spin-ladder model and
identify the corresponding
discrete symmetry with that of the related Ising models.

Since spin-rotational invariance remains
unbroken,
the string order
parameter  must respect this symmetry. However,  Abelian bosonization
is not an explicitly SU(2) invariant procedure. For this
reason, it turns out that it is the $z$-component of the string operator that
acquires a simple form in the continuum limit. On the other hand, due to the
unbroken SU(2) symmetry, the very choice
of the quantization ($z$-) axis is arbitrary; therefore the expectation values
for all components of the string operator will coincide.

To construct a string order parameter $O^z (n,m)$
 for the spin-ladder model, we shall follow the same route as that previously
used for the bond-alternating S = 1/2 chain \cite{Kohmoto} (technical details
are given
in the Appendix B.).  We start from the lattice version of the model,
construct a product of two spin-1/2 operators belonging to the $j$-th rung,
$S^z _1 (j) S^z _2 (j)$, and then take a product over all rungs between
$j = n$ and $j = m$:
\begin{equation}
O^z (n,m) = \prod_{j=n}^{m} (- 4 S^z _1 (j) S^z _2 (j))
= \exp \left(  \mbox{i} \pi \sum_{j=n}^{m} [~S^z _1 (j) + S^z _2 (j)~]\right)
\end{equation}
Assuming that $|m - n| \gg 1$, we pass to the continuum limit in the
exponential and retain only the smooth parts of the spin operators
expressing them in terms of the spin currents $J_{a; R,L}^{z} (x),~(a = 1,2)$:
\begin{eqnarray}
O^z (x, y) &=& \exp \left( \pm \mbox{i}\pi \sum_{a=1,2} \int_{x}^{y} dx'
S^z _a (x') \right) \nonumber\\
&=&  \exp \left( \pm \mbox{i}\pi \sum_{a=1,2} \int_{x}^{y} dx'
[J_{a; R}^{z} (x') + J_{a; R,L}^{z} (x')] \right).
\end{eqnarray}
Using Eqs.(\ref{+-defs}),(\ref{eq:bos-z}), we find that the
exponential is expressed in terms of the field $\phi_{+}$ only.
 Thus we find a very transparent representation for the
string operator:
\begin{equation}
O^z (x, y) = \exp ( \mbox{i} \sqrt{\pi} [\phi_{+} (x) - \phi_{+} (y)] )
\end{equation}

Using Eq.(\ref{eq:sig}),
\begin{equation}
\exp ( i \sqrt{\pi} \phi_{+} (x) ) \sim \sigma_1 \sigma_2 + i \mu_1 \mu_2
\end{equation}
we find that the string operator is expressed in terms of the Ising order
and disorder operators. For either sign of $J_{\perp}$, we find that, in the
limit $|x - x'| \rightarrow \infty$, the vacuum expectation value
of $O^z (x, y)$ is indeed nonzero:
\begin{eqnarray}
\lim_{|x - x'| \rightarrow \infty} \langle O^z (x, y)\rangle
&\sim& <\sigma_1>^2 ~ <\sigma_2>^2 ~=~ <\sigma>^4 \neq 0, ~~~J_{\perp} < 0
\label{eq:fin1}\\
\lim_{|x - x'| \rightarrow \infty} \langle O^z (x, y)\rangle
&\sim& <\mu_1>^2 ~ <\mu_2>^2 ~=~ <\mu>^4 \neq 0, ~~~J_{\perp} > 0
\label{eq:fin2}
\end{eqnarray}

As in the case of the bond-alternating spin chain, the nonvanishing expectation
value of the string order parameter in the limit of infinite string manifests
breakdown of a discrete $Z_2 \times Z_2$ symmetry. This is the symmetry of two
decoupled Ising models described by the Hamiltonian $H_+$ in the Majorana
fermion representation (\ref{eq:M+}): $H_+ = H_m [\xi^1] + H_m [\xi^2]$ remains
invariant with respect to sign inversion of both chiral
components of each Majorana spinor,
$
\xi^a _{R,L} \rightarrow - \xi^a _{R,L}, ~(a = 1,2.)
$
Under these transformations, the Ising order and disorder parameters change
their signs. On the other hand, since the two Majorana fermions are massive,
this symmetry is broken
in the {\it ground state} of $H_+$: the mass terms break the duality symmetry
$\xi^a _L \rightarrow - \xi^a _L$, $\xi^a _R \rightarrow \xi^a _R$. This
amounts to finite expectation values of the Ising variables $\sigma_1$ and
$\sigma_2$ (or $\mu_1$ and $\mu_2$),
which in turn results in a nonzero string order parameter, as
shown in Eqs.(\ref{eq:fin1}) and (\ref{eq:fin2}).

\section{Conclusions}

 As the reader might see the spin ladder presents an exciting
opportunity to study  a formation of massive spin S = 1 and S = 0 particles
which appear as bound states of spin  S = 1/2 excitations of
individual Heisenberg chains. At small interchain coupling $|J_{\perp}|
<< J_{||}$ the  masses  of these  particles
are of order of $|J_{\perp}|$. The S = 1 branch is always lower
independently of the sign of $J_{\perp}$. At  $J_{\perp}/J_{||}
\rightarrow 0$ the singlet spectral gap is three times as large as the
triplet one. The comparative smallness
of the interchain coupling
creates room for the crossover from the single chain behaviour  to the
strong coupling regime which resembles the S = 1 chain.
The imaginary part  of the
dynamical spin susceptibility $\chi''(\omega, q; q_{\perp})$
calculated in Section 3 contains essential information about this
crossover. At small energies the susceptibility  exhibits a
sharp peak around $q = \pi$
correspondinding to the stable S = 1 massive particle; at
energies  $\omega > 3m$ $\chi''(\omega, q)$ exhibits an  incoherent
tail originating from the  multi-particle processes. Below the
$5m$-threshould  the singlet branch does not contribute to
$\chi''(\omega, q)$ and the latter
coincides with the susceptibility of a S = 1 chain. The  contribution
from the singlet mode becomes essential at energies much greater
than the spectral  gap
and  the susceptibility asymptotically approaches  its value for
a spin-1/2 chain. We emphasise that the
described picture holds only in the ideal limit $J_{\perp}/J_{||}
\rightarrow 0$. We suppose that in real systems it will be difficult
to make this ratio less than 0.1.

\section{Acknowledgements}

We acknowledge useful discussions with I.E.Dzyaloshinskii and A.M.Finkel'stein.
A. A. N. is grateful to the Department of Theoretical Physics of
University of Oxford for kind hospitality and acknowledges the support
from the ESRSC grant No. GR/K4 1229.

\newpage
\appendix
\section*{Appendix A: Basic facts about bosonization}

Antiferromagnetic spin chain Hamiltonians, such as the Heisenberg Hamiltonian

\begin{equation}
H = J \sum_{n=1}^{N} {\bf S}_n \cdot {\bf S}_{n+1},
{}~~~(S = 1/2, ~~J > 0) \label{eq:single}
\end{equation}
can be mapped onto fermionic theories. Using bosonization,
 these can be recast as generalised Sine-Gordon or WZW
models. This is useful because a great deal is known about these
theories, such as correlation functions, scaling dimensions of
operators etc. A brief summary of this approach is given below.

Following Refs. \cite{aff}, we start from  a {\it symmetry preserving}
fermionization of the spin operators
\begin{equation}
{\bf S}_n = \psi^{\dagger}_{n \alpha} \frac{\vec{\sigma}_{\alpha \beta}}{2}
\psi_{n \beta} \label{eq:fermion}
\end{equation}
To eliminate the redundant zero- and double-occupancy states, the constraint
$ \sum_{\alpha} \psi^{\dagger}_{n \alpha} \psi_{n \alpha} = 1$ for all
lattice sites $n$ should be imposed. Such a constraint will effectively work,
if one considers a 1/2-filled, $U > 0$ Hubbard model for the field $\psi_{n
\alpha}$. In this model, a Mott-Hubbard charge gap $m_c$ is known to exist for
{\it any} positive $U$. Therefore, at low energies, $|E| \ll m_c$, only spin
excitations remain;
those describe universal dynamical properties of the spin-chain model
(\ref{eq:single}) in the continuum limit.

Assuming that $U \ll t$, we linearize the free-particle spectrum near two
Fermi points, $\pm k_F~ (k_F = \pi /2 a_0)$, and decompose the Fermi field
into right-moving and left-moving chiral components:
\begin{equation}
\psi_{n \alpha} \rightarrow \sqrt{a_0}~\psi_{\alpha} (x), ~~~
\psi_{\alpha} (x) = (- \mbox{i})^n \psi_{R \alpha} (x) + \mbox{i}^n \psi_{L
\alpha} (x)
\end{equation}
We then introduce the scalar [U(1)] and vector [SU(2)] currents
(the local charge and spin densities)

\begin{equation}
J_{R,L} = ~:\psi^{\dagger}_{R,L; \alpha} \psi_{R,L; \beta}: , ~~~~~
\\
{\bf J}_{R,L} = ~:\psi^{\dagger}_{R,L; \alpha} \frac{\vec{\sigma}_{\alpha
\beta}}{2} \psi_{R,L; \beta}: \label{eq:SU(2)curr}
\end{equation}
satisfying anomalous (U(1) and SU(2)) Kac-Moody algebras:
\begin{equation}
[J_R (x), J_R (x')] =
\frac{1}{\mbox{i} \pi} \delta' (x- x'), \label{eq:U(1)KM}
\end{equation}
\begin{equation}
[J^a _R(x), J^b _R(x')] = \mbox{i} \epsilon^{abc} J^c_{R} (x) \delta (x - x') -
\frac{\mbox{i}}{4 \pi} \delta^{ab} \delta' (x- x') \label{eq:SU(2)KM}
\end{equation}
(with similar relations for the left components). These algebras lead to
fermion-boson duality which allows to represent the Hamiltonian of free
fermions
as a sum of two independent (commuting) contributions of gapless charge and
spin collective modes (Sugawara form):
\begin{eqnarray}
H^0 = H^0_{U(1)} + H^0_{SU(2)} \nonumber\\
= \int dx [~\frac{\pi v_F}{2} ~(:J_R J_R: + :J_L J_L:)
+
\frac{2 \pi v_F}{3}  ~(:\bf{J}_R \cdot \bf{J}_R: + :\bf{J}_L \cdot \bf{J}_L:)]
\label{eq:csHam}
\end{eqnarray}
The charge part is equivalently described in terms of a massless scalar field
$\phi_c$.
Under identifications
$
J_R + J_L = \frac{1}{\sqrt{\pi}} \partial_x \phi_c,$
$
J_R - J_L = - \frac{1}{\sqrt{\pi}} \Pi_c$,
where $\Pi_c$ is the momentum conjugate to the field $\phi_c$, one obtains
\begin{equation}
H^0_{U(1)} = \frac{v_s}{2} \int dx ~[ \Pi^2 _c(x) + (\partial_x \phi_c (x))^2 ]
\label{eq:Hboson-c}
\end{equation}
The spin part $H^0_{SU(2)}$ represents the level $k = 1$ $SU(2)$-symmetric
critical Wess-Zumino-Witten (WZW) model.

A weak Hubbard interaction preserves the important property of charge-spin
separation, $H_{Hubbard} = H_c + H_s$.
Umklapp processes relevant at 1/2-filling transform $H^0_{U(1)}$ to  a quantum
sine-Gordon model
\begin{equation}
H_c = \int dx ~[ \frac{v_c}{2} \left( \Pi^2 _c + (\partial_x \phi_c)^2 \right)
+ const ~g~ \cos \beta_c \phi_c ] \label{eq:SG}
\end{equation}
which at $g \sim U/t > 0$ occurs in its strong-coupling, massive phase
($\beta^2 < 8 \pi$), with the single-soliton mass $m_c$ being just the
the Mott-Hubbard commensurability gap.

In the spin  sector, interaction $- 2 g {\bf J}_R \cdot{\bf J}_L$ is added
to $H^0 _{SU(2)}$.
This interaction is marginally {\it
irrelevant} (since $g > 0$). Therefore, the
universal scaling properties of the Heisenberg S = 1/2 spin chain
(\ref{eq:single}) are described by the level $k = 1$ WZW model $H^0_{SU(2)}$
\cite{aff}.

The possibility of an Abelian bosonization of the Heisenberg chain
(\ref{eq:single}) stems from the fact that conformal charges of the $k = 1$
$SU(2)$ WZW models and free massless Bose field coincide:
$ C^{WZW}_{SU(2), k=1} = C_{boson} = 1$. Using relations
$ (1/3)~ {\bf J}_{R(L)} \cdot {\bf J}_{R(L)} = J^z _{R(L)} J^z _{R(L)}$,
$H^0_{SU(2)}$ can be expressed in terms of $J^z$-currents only;
introducing then a pair of canonical variables,
$\phi_s$ and $\Pi_s$, via
\begin{equation}
J^z _R + J^z _L = \frac{1}{\sqrt{2 \pi}} \partial_x \phi_s,~~~
J^z _R - J^z _L = - \frac{1}{\sqrt{2 \pi}} \Pi_s, \label{eq:bos-z}
\end{equation}
one finds
\begin{equation}
H^0_{SU(2)} \rightarrow  H_B = \frac{v_s}{2} \int dx ~[ \Pi^2 _s (x) +
(\partial_x \phi_s (x))^2 ]
\label{eq:Hboson-s}
\end{equation}

The price we pay for this simplification is the loss of spin rotational
invariance in the bosonized structure of the spin currents:
the $J^x$ and $J^y$ cannot be represented as simply as $J^z$, and require
bosonization of the Fermi fields:
\begin{equation}
\psi_{R,L; \alpha} (x) \simeq (2 \pi a_0)^{-1/2}
\exp \left( \pm i \sqrt{4 \pi}~ \varphi_{R,L; \alpha} (x)\right)
\label{eq:bosoniz}
\end{equation}

Linear combinations
$$
\Phi_{\alpha} = \varphi_{R \alpha} + \varphi_{L \alpha}, ~~~~
\Theta_{\alpha} = - \varphi_{R \alpha} + \varphi_{L \alpha}
$$
constitute scalar fields $\Phi_{\alpha}$ and their dual counterparts
$\Theta_{\alpha}$ introduced for each spin component. The fields describing the
charge and spin degrees of freedom are defined as follows:

\begin{eqnarray}
\phi_c &=& \frac{\Phi_{\uparrow} + \Phi_{\downarrow}}{\sqrt{2}},~~~~
\theta_c = \frac{\Theta_{\uparrow} + \Theta_{\downarrow}}{\sqrt{2}} \nonumber\\
\phi_s &=& \frac{\Phi_{\uparrow} - \Phi_{\downarrow}}{\sqrt{2}},~~~~
\theta_s = \frac{\Theta_{\uparrow} - \Theta_{\downarrow}}{\sqrt{2}}
\end{eqnarray}
where $\partial_x \theta_{c,s} = \Pi_{c,s}$.

To bosonize $J^{\pm}_{R,L}$, use (\ref{eq:bosoniz}) to obtain:
\begin{eqnarray}
J^+ _R &=& \psi^{\dagger} _{R \uparrow} \psi_{R \downarrow}
= \frac{1}{2 \pi a_0} \exp (- i \sqrt{2 \pi} (\phi_s - \theta_s ) ) \nonumber\\
J^+ _L &=& \psi^{\dagger} _{L \uparrow} \psi_{L \downarrow}
= \frac{1}{2 \pi a_0} \exp (i \sqrt{2 \pi} (\phi_s +\theta_s ) )
\label{eq:bos-pm}
\end{eqnarray}
Note that, as expected, the charge field $\phi_c$ does not contribute to the
spin SU(2) currents. Moreover, despite the fact that the definitions
(\ref{eq:bos-pm})
contain cutoff $a_0$ explicitly, the current-current correlation
functions are cutoff independent and reveal the underlying SU(2)
symmetry:
\begin{equation}
< J^a (x) J^b (x') > = - \frac{\delta^{ab}}{4 \pi^2} \frac{1}{(x - x')^2}
\end{equation}

The SU(2) currents ${\bf J}_R (x), {\bf J}_L (x)$ determine the smooth parts
of the spin operators in the continuum limit. Namely, at $a_0 \rightarrow 0$
\begin{equation}
{\bf S}_n \rightarrow a_0 {\bf S} (x),~~~
{\bf S} (x) = {\bf J}_R (x) + {\bf J}_L (x) + (-1)^n {\bf n} (x)
\label{eq:Scont}
\end{equation}
where
\begin{equation}
{\bf n} (x) = \psi^{\dagger}_{R \alpha} (x) \frac{\vec{\sigma}_{\alpha
\beta}}{2} \psi_{L \beta} (x)  + h.c. \label{eq:stag}
\end{equation}
is the staggered part of the local spin density.

When bosonizing (\ref{eq:stag}), the (redundant) charge excitations emerge,
since off-diagonal
bilinears like $\psi^{\dagger}_R \psi_L$ and $\psi^{\dagger}_L \psi_R$ describe
particle-hole {\it charge} excitations with momentum transfer $\pm 2 k_F$.
We find:
\begin{eqnarray*}
n^z &=&  - \frac{1}{\pi a_0} \sin \sqrt{2 \pi} \phi_c ~\cos \sqrt{2 \pi} \phi_s
\nonumber\\
n^\pm &=& - \frac{1}{\pi a_0} \sin (\sqrt{2 \pi} \phi_c / 2)~ \exp ( \pm i
\sqrt{2 \pi} \theta_s)
\end{eqnarray*}
Being interested in the energy range $|E| \ll m_c$, one can replace
 the charge operator $\sin ( \sqrt{2 \pi}\phi_c / 2)$ by its
 nonzero vacuum expectation value; we denote this (nonuniversal) value by
$
\lambda = - < \sin (\sqrt{2 \pi} \phi_c / 2) >
$
and arrive at bosonization formulas for ${\bf n} (x)$:
\begin{eqnarray}
n^z (x) &=& \frac{\lambda}{\pi a_0} \cos \sqrt{2 \pi} \phi_s (x) \nonumber\\
n^\pm (x) &=& \frac{\lambda}{\pi a_0} \exp [ \pm i \sqrt{2 \pi} \theta_s (x)]
\label{eq:bosstag}
\end{eqnarray}

This completes the bosonization of the spin operators for the
isotropic Heisenberg chain. Notice that the critical dimensions of the smooth
and staggered parts of
the spin densities are different:
\begin{equation}
dim~J^a = 1, ~~~ dim~n^a = 1/2. \label{eq:scaldim}
\end{equation}

\section*{Appendix B. Hidden $Z_2 \times Z_2$ symmetry and string order
parameter
in the bond-alternating S = 1/2 Heisenberg chain}

In addition to the S = 1/2 spin-ladder model, there is another
system  which is related to the S = 1 spin chain - the spin-1/2 chain with
alternating ferromagnetic and antiferromagnetic bonds:

\begin{equation}
H = 4J \sum_{j=1}^{N/2} [( {\bf S}_{2j-1}\cdot {\bf S}_{2j} )
- \beta ({\bf S}_{2j} \cdot  {\bf S}_{2j+1}] \label{eq:alt}
\end{equation}
This model is instructive in the sense that
the string order parameter, whose nonzero expectation value
signals breakdown of a hidden discrete symmetry, can be easily constructed
\cite{Kohmoto}.
The analogous construction is then directly generalized for the
spin-ladder model.

A gap in the excitation spectrum of the model (\ref{eq:alt}) persists in the
whole range $0 < \beta < \infty$. At $\beta = 0$ the ground state of
model represents an array of disconnected singlets.
At $\beta \gg 1$, strong ferromagnetic coupling between the
spins on the $<2j, 2j+1>$-bonds leads to the formation of local triplets,
and the model (\ref{eq:alt}) reduces to a S = 1 Heisenberg chain. Using
a nonlocal unitary transformation, Kohmoto and Tasaki\cite{Kohmoto}
 have demonstrated equivalence of the model
(\ref{eq:alt}) to a system of two coupled quantum Ising chains, {\it i.e.}
two coupled 2d Ising models. This transformation provides an exact
representation of the spin
operators $S^{\alpha} _n$ as products of two Ising-like order
$( \sigma, \tau )$ and disorder $( \tilde{\sigma}, \tilde{\tau} )$ operators,
essentially a lattice version of relations (\ref{eq:sig}), (\ref{eq:siga})
(see, e.g., \cite{nersesyan}).
Nearest-neighbor bilinears of the original spin operators take the form
\begin{eqnarray}
4 S^{x}_{2j} S^{x}_{2j+1} = - \sigma^{z}_j \sigma^{z}_{j+1}, &&
4 S^{x}_{2j-1} S^{x}_{2j} = - \tau^{x}_j\nonumber\\
4 S^{y}_{2j} S^{y}_{2j+1} = - \tau^{z}_j \tau^{z}_{j+1}, &&
4 S^{y}_{2j-1} S^{y}_{2j} = - \sigma^{x}_j\nonumber\\
4 S^{z}_{2j} S^{z}_{2j+1} = - \sigma^{z}_j \sigma^{z}_{j+1} \tau^{z}_j
\tau^{z}_{j+1}, &&
4 S^{z}_{2j-1} S^{z}_{2j} = - \sigma^{x}_{j} \tau^{x}_j
\label{eq:prod}
\end{eqnarray}
where
\begin{eqnarray}
\sigma^{x}_{j} = \tilde{\sigma}^{z}_{j-1/2}~ \tilde{\sigma}^{z}_{j+1/2},&&
\tau^{x}_{j} = \tilde{\tau}^{z}_{j-1/2}~ \tilde{\tau}^{z}_{j+1/2}
\label{eq:dualtr}  \\
\tilde{\sigma}^{z}_{j+1/2} = \prod_{l=j+1}^{N/2} \sigma^{x}_{l}, &&
 \tilde{\tau}^{z}_{j+1/2} = \prod_{l=1}^{j-1} \tau^{x}_{l}
\end{eqnarray}

Relations (\ref{eq:prod}) make the Hamiltonian  (\ref{eq:alt}) equivalent to
two coupled quantum Ising chains:
\begin{equation}
H = -J \sum_{j=1}^{N/2} [ ~( \beta \sigma^z _{j} \sigma^z _{j+1} + \sigma^x _j)
+ ( \beta \tau^z _{j} \tau^z _{j+1} + \tau^x _j)
 + ( \beta \sigma^z _{j} \sigma^z _{j+1} \tau^z _{j} \tau^z _{j+1}
+ \sigma^x _j \tau^x _j) ~] \label{2Is.ch.}
\end{equation}
The model (\ref{2Is.ch.}) is invariant under independent rotations of the
$\sigma$ and $\tau$ spins by angle $\pi$ about the spin $x$-axis
which comprise a $Z_2 \times Z_2$ group. Since this group is discrete, it
can be spontaneously broken, in which case the spectrum of the system would be
massive. It is easily understood from (\ref{2Is.ch.}) that, in the limit of
large positive $\beta$
when the model reduces to the $S = 1$ chain, the $Z_2 \times Z_2$-symmetry
is broken, with
\begin{equation}
<\sigma^{z}_{j}> ~=~ <\tau^{z}_{j}> ~=~ <\sigma^{z}_{j} \tau^{z}_{j}> ~\neq ~0,
\label{eq:order}
\end{equation}
(It has been used in Eq.(\ref{eq:order}) that, under
transformation $\mu^z _j = \sigma^z _j \tau^z _j$ to a new pair of variables,
$\mu^z _j$ and $\tau^z _j$, the two-chain Hamiltonian (\ref{2Is.ch.})
preserves its form.).

Representation (\ref{eq:prod})
hints to the way how an order parameter measuring breakdown of the $Z_2 \times
Z_2$-symmetry  should be constructed out of the spin
operators $S^{\alpha}_{n}$.
Following Kohmoto and Tasaki, consider a product
\begin{eqnarray}
\prod_{l=2k}^{2n-1} 2 S^x _l = \prod_{j=k}^{n-1} 4 S^x _{2j} S^x _{2j+1}
= \prod_{j=k}^{n-1} ( - \sigma^z _{j} \sigma^z _{j+1})\nonumber\\
= (-1)^{n-k} (\sigma^z _{k} \sigma^z _{k+1})
 (\sigma^z _{k+1} \sigma^z _{k+2}) \cdots
(\sigma^z _{n-1} \sigma^z _{n}) = (-1)^{n-k} \sigma^z _k \sigma^z _n
\end{eqnarray}
Using the relation $i \sigma^{\alpha}_{j} = \exp (i \pi \sigma^{\alpha}_{j} /
2)$,
we find that
\begin{equation}
O^x (k,n) \equiv \exp \left ( i \pi \sum_{l=2k}^{2n-1} S^x _l \right )
= \sigma^z _k \sigma^z _n . \label{eq:str-x}
\end{equation}
This is the $x$-component of the string order operator. According to
(\ref{eq:order}), in the limit $|k - n| \rightarrow \infty$,
its vacuum expectation value is nonzero:
\begin{equation}
< O^x (k,n) > ~\rightarrow~ <\sigma>^{2} \neq 0 \label{eq:str-exp}
\end{equation}

It is important that the string always contains an even number of sites,
starting at an even site and ending at an odd site.
For a string starting at an odd site and ending at an even site,
the corresponding string operator is expressed in
terms of disorder operators and therefore has zero expectation value:
\begin{eqnarray*}
\prod_{l=2k+1}^{2n} 2 S^x _l = \prod_{j=k+1}^{n} 4 S^x _{2j-1} S^x _{2j}
= (-1)^{n-k} (\tilde{\tau}^z _{k+1/2} \tilde{\tau}^z _{k+3/2})
(\tilde{\tau}^z _{k+3/2} \tilde{\tau}^z _{k+5/2}) \cdots
(\tilde{\tau}^z _{n-1/2} \tilde{\tau}^z _{n+1/2}) \\
 = (-1)^{n-k} \tilde{\tau}^z _{k+1/2} \tilde{\tau}^z _{n+1/2}
\end{eqnarray*}
The $y$ and $z$ components of the string operator are constructed in a similar
manner:
\begin{eqnarray}
O^y (k,n) = \exp \left( \mbox{i} \pi \sum_{l=2k}^{2n-1} S^y _l \right)
= \prod_{l=2k}^{2n-1} 2 \mbox{i} S^y _l
= \prod_{j=k}^{n-1} (- 4 S^y _{2j} S^y _{2j+1}) \nonumber\\
= \prod_{j=k}^{n-1} \tau^z _{j} \tau^z_{j+1} = \tau^z _k \tau^z _n
\end{eqnarray}
\begin{equation}
O^z (k,n) = \exp \left( \mbox{i} \pi \sum_{l=2k}^{2n-1} S^z _l \right)
= \sigma^z _{k} \tau^z _k \sigma^z_{n} \tau^z _{n}
\end{equation}
The SU(2) invariance of the expectation value of the string order parameter
\begin{equation}
O^{\alpha} (k,n) = \exp \left( \mbox{i} \pi \sum_{l=2k}^{2n-1} S^{\alpha} _l
\right),~~
(S = 1/2,~~ \alpha = x,y,z) \label{eq:strSU(2)}
\end{equation}
follows from (\ref{eq:order}).

Notice that
in the limiting case $\beta \gg 1$, the string order parameter
(\ref{eq:strSU(2)}) for the S = 1/2 bond-alternating chain automatically
transforms to the exponential of the string order parameter
(\ref{eq:strS=1}) for the S = 1 chain.

\end{document}